\documentclass[a4paper,12pt]{article}

\usepackage[american]{babel}
\usepackage{amsmath,amssymb,amsthm}

\theoremstyle{plain}
\newtheorem{theo}{Theorem}
\newtheorem{rema}{Remark}
\newtheorem{exam}{Example}

\newcommand\N{{\mathbb{N}}}
\newcommand\R{{\mathbb{R}}}
\newcommand\C{{\mathbb{C}}}
\renewcommand{\i}{{\rm i}}
\newcommand{\x}{\mathbf{x}}
\renewcommand{\k}{\mathbf{k}}
\newcommand{\z}{\mathbf{z}}
\newcommand{\w}{\mathbf{w}}
\renewcommand{\d}{{\rm d}}

\renewcommand{\Im}{{\rm Im}}

\begin{document}

\title{On the nonlocality of the state and wave equation of Treeby and Cox}

\author{ Richard Kowar\\
Department of Mathematics, University of Innsbruck, \\
Technikerstrasse 21a, A-6020, Innsbruck, Austria}

\maketitle

\begin{abstract}
In this paper it is shown that the state equation of Treeby and Cox [B. E. Treeby and B. T. Cox, J. Acoust. Soc. Am. 
\textbf{127} 5, (2010)] is \emph{nonlocal}, more precisely, a \emph{local} density variation causes an \emph{instant global} 
pressure variation and a \emph{local} pressure variation can only be caused by an \emph{instant global} density variation.
This is in contrast to all frequency dependent dissipative state equations known to the author.    
Moreover, it is shown that the Green function $G$ of the wave equation of Treeby and Cox cannot have a \emph{finite} wave 
front speed, i.e. there exists no finite $c_F>0$ such that 
$$
  G(\x,t) = 0 \qquad\mbox{for}\qquad  |\x|/c_F > t
$$
holds, where $|\x|/c_F$ corresponds to the \emph{travel time} of a wave propagating with speed $c_F$ from point $\mathbf{0}$ 
to point $\x$. As a consequence, the density and pressure waves satisfying (i) the state equation of Treeby 
and Cox, (ii) the equation of motion and (iii) the equation of continuity do not have a \emph{finite} wave front speed. 
\end{abstract}

\section{Introduction}\label{sec-intro}

By the authors of~\cite{TrZhCo10}, the \emph{pressure state equation}
\begin{equation}\label{stateeqTC}
   p =   c_0^2\,\rho
       - a_0 \,  (-\Delta)^{\gamma-1}\,\frac{\partial \rho}{\partial t} 
       - b_0\, (-\Delta)^{\gamma-1/2}\,\rho
\end{equation}
with constants $\gamma\in (0,\frac{1}{2})\cup (\frac{1}{2},\frac{3}{2})$, $c_0>0$, $a_0(\gamma)>0$ and 
$b_0=-a_0\,c_0\,\tan(\pi\,\gamma)$ 
was postulated for some dissipative media arising in \emph{biological} and \emph{medical applications}. 
Based on this \emph{state equation}, they also derived the pressure wave equation
\begin{equation}\label{waveeqTC}
    \frac{\partial^2 p}{\partial t^2}
          + a_0\,  (-\Delta)^{\gamma}\,\frac{\partial p}{\partial t}
          + b_0\, (-\Delta)^{\gamma+1/2}\,p
          - c_0^2\,\Delta p
             = 0 \,,
\end{equation}
which is an extension of the \emph{wave equation of Chen and Holm} modeled by the authors of~\cite{ChHo04}. 
That the latter wave equation does not have a finite wave front speed was shown by the authors of~\cite{KoSc12}. 
In this paper, we investigate the following two issues arising from the model of Treeby and Cox for 
\begin{equation}\label{Assabc}
     c_0>0\,,\quad a_0>0\,,\quad b_0\in\R\quad   \mbox{(all independent of $\gamma$)}\quad \mbox{and}\quad \gamma > 0\,. 
\end{equation}

\subsection*{a) Nonlocality of the state equation}

The state equation of \emph{standard dissipative} pressure waves has the form
\begin{equation}\label{stateeqst}
    p(\x,t) = c_0^2\, \int_0^t K_1(t-s)\,\rho(\x,s)\,\d s
    \qquad\quad  (\x\in\R^3,\,t\in\R) 
\end{equation} 
for any density variation initiated at time $t=0$ and kernel $K_1$ satisfying 
\begin{equation}\label{KKrel}
    K_1(t) = 0 \qquad \mbox{for} \qquad t<0  \qquad\quad \mbox{(a \emph{causality condition}).}
\end{equation} 
%Here $K_1$ depends on the (complex) attenuation law. 
This state equation is local and consequently \emph{if} $\rho$ is a density wave with finite wave front speed, then the 
pressure wave satisfying the state equation~(\ref{stateeqst}) has a finite wave front speed, too.

In contrast, we show in this paper that the \emph{state equation of Treeby and Cox} is \emph{nonlocal} for $\gamma>0$, i.e. 
\begin{itemize}
\item a \emph{local} density variation $\delta \rho$ causes an \emph{instant global} pressure variation $\delta p$ and

\item a \emph{local} pressure variation $\delta p$ can only be caused by an \emph{instant global} density variation $\delta \rho$.

\end{itemize} 
As a consequence, if $\rho$ denotes a density wave with finite wave front speed, then the pressure wave defined by the 
state equation of Treeby and Cox does not have a finite wave front speed. 
Obviously, this behavior violates \emph{cause} and \emph{effect}, at least in the framework of classical physics. 
In addition, this nonlocal behavior is in contrast to all frequency dependent dissipative state equations known to the 
author (cf.~\cite{Sza94,Sza95,WatHugBraMil00,KoSc12}).

\subsection*{b) Infinite speed of the wave front}

The state equation not only relates the pressure with the density, but in combination with the \emph{equation of motion} and 
\emph{equation of continuity}, it implies a density wave equation. As a consequence, it determines the causal behavior of the 
respective density wave, too. 
We show that the density wave as well as the pressure wave implied by the Treeby and Cox model do not have a finite wave front 
speed $c_F$ for $\gamma>0$. More precisely, if $G$ denotes the \emph{Green function} of the density wave equation, then for 
every $c_F>0$ 
\begin{equation}\label{causcondT}
        G(\x,t) = 0 \qquad\mbox{for}\qquad  |\x|/c_F > t
\end{equation}
does not hold. Here $|\x|/c_F$ corresponds to the \emph{travel time} of a wave front traveling with speed $c_F$ 
from point  $\mathbf{0}$ to point $\x$. 
The physical interpretation of Eq.~(\ref{causcondT}) is as follows: Consider a dissipative spherical wave, 
i.e. $G$, generated in location $\mathbf{0}$ at time $0$. Then an observer in location $\x\in\R^3\backslash\{\mathbf{0}\}$ 
feels the wave not before the travel time $\frac{|\x|}{c_F}$ had passed. If he feels the wave sooner the wave front speed 
is larger than $c_F$ and otherwise it is smaller. However, if he feels it for every arbitrary small time period, then 
$c_F=\infty$. 

Because the wave front speed of a non-dissipative wave is finite, it seems not reasonable that a dissipative wave 
model has an infinite wave front speed.

\begin{rema}
If we consider the "reversed" state equation 
$$
     \rho(\x,t)=c_0^{-2}\, \int_0^t K_2(t-s)\,p(\x,s)\,\d s 
$$ 
with $(K_2 *_t K_1)(t) = \delta(t)$, 
then condition~(\ref{KKrel}) with $K_1$ replaced by $K_2$ corresponds to the \emph{Kramer-Kronig relations}
which are satisfied for all standard frequency dependent dissipation models known to the author 
(cf.~\cite{Sza94,Sza95,WatHugBraMil00,KoSc12}). As above, it follows that 
the Kramer-Kronig relations imply a finite wave front speed for the wave $\rho$ (output of linear system), \emph{if} the wave 
$p$ (input) has a finite wave front speed.\footnote{This is an interpretation of the \emph{Kramer-Kronig relation} for the 
case $K_2=K_2(t)$.} However, it does not follow that the pressure wave $p$ (input) satisfying, in addition, the 
\emph{equation of motion} and \emph{equation of continuity} has a finite wave front speed. This is partly a 
misapprehension in the literature. For more detail, we refer to~\cite{KoScBo11,KoSc12}.   
\end{rema}

This paper was written with the intention to keep its contents as short and simple as possible such that it is readable 
for a broad range of physicists and applied mathematicians. 
It is organized as follows. First, we investigate and discuss the properties of state equations of dissipative media. 
Second, we show that the density wave and the pressure wave of the Treeby and Cox model do not have a finite wave 
front speed. For the convenience of the reader, we put the mathematical part of this paper in a self-contained appendix.

\section{Properties of the state equation}
\label{sec-stateeq}

In this section we show the nonlocality of the state equation~(\ref{stateeqTC}) of Treeby and Cox. 
For a better understanding, we shortly discuss the non-dissipative and the frequency dissipative case. 
In what follows, $\delta(\x)$ and $\delta(t)$ denote the \emph{dirac distribution} on $\R^3$ and $\R$, respectively.

\subsection*{Dissipation-free case}

For a non-viscous fluid the state equation can be modeled by 
$$
          p = c_0^2\,\rho\,,
$$
which is nothing else but Eq.~(\ref{stateeqTC}) for $\gamma=0$ with $a_0=0$ and $b_0=0$. As a consequence, we have 
$$
       \delta \rho(\x,t) = \delta(\x-\x_0)\,\delta(t)
                    \qquad \Leftrightarrow \qquad
       \delta p(\x,t) = c_0^2\,\delta(\x-\x_0)\,\delta(t) \,, 
$$
i.e. a \emph{local} density variation $\delta \rho$ in point $\x_0$ at time $t=0$ causes a \emph{local} pressure variation 
$\delta p$ in the same point at the same time and vice versa. Due to this local behavior in space and time, $p$ has a 
finite wave front speed if and only if $\rho$ has a finite wave front speed.

\subsection*{Standard frequency dependent case}

For the case of standard frequency dependent dissipation, the state equation 
$$
          p = c_0^2\, (K_1 *_t \rho)   \qquad\quad\mbox{($*_t$ time convolution)}\,
$$
satisfying the causality condition~(\ref{KKrel}) implies
$$
       \delta \rho(\x,t) = \delta(\x-\x_0)\,\delta(t)
                \qquad \Rightarrow \qquad
       \delta p(\x,t) = c_0^2\,K_1(t)\,\delta(\x-\x_0)  \,,
$$
i.e. a \emph{local} density variation $\delta \rho$ in point $\x_0$ at time $t=0$ causes a \emph{local} pressure variation 
$\delta p$ in the same point for the time points $t\geq 0$. 
We still have a local behavior in space, but also a memory effect modeled by the function $t\mapsto K_1(t)$. 
Due to $K_1=K_1(t)$ and causality condition~(\ref{KKrel}), $p$ has a finite wave front speed, \emph{if} 
$\rho$ has a finite wave front speed. Mathematically, this means that 
the statement 
$$
        \rho(\x,t) = 0 \quad\mbox{for}\quad  \frac{|\x|}{c_F} > t    \qquad\Rightarrow\qquad 
        (K_1 *_t \rho)(\x,t) = 0 \quad\mbox{for}\quad  \frac{|\x|}{c_F} > t
$$
holds. 
%We emphasize that the statement $A \Rightarrow B$ is true does not mean that statement $A$ is true. 

\subsection*{Model case of Treeby and Cox}

The following statement concerning the distribution 
\begin{equation}\label{defA}
         A^\gamma(\x) :=  (-\Delta)^\gamma\,\delta(\x)     
         \qquad \mbox{for $\gamma>0$ with $\gamma\not\in\{1,\,2,\,3,\,\ldots\}$} 
\end{equation}
is shown in the appendix (cf. Example~\ref{exam:PW01}): 
\begin{itemize}
\item [(A)] For every $n\in\N$ there exists an $\x_n\in\R^3$ with $|\x_n|\geq n$ such that\footnote{Notice that $|A(\x_n)|$ 
            may be small for large $n$, but it does not vanish.}
$$
         A^\gamma(\x_n)   \qquad \mbox{does not vanish}
$$ 
\end{itemize} 
for $\gamma>0$ with $\gamma\not\in\{1,\,2,\,3,\,\ldots\}$. 
In particular, this means that $(-\Delta)^\gamma$ is not a \emph{local operator} for $\gamma>0$ with $\gamma\not\in\N$. \\

From the state equation~(\ref{stateeqTC}) for $\gamma>0$, it follows that
$$
       \delta \rho(\x,t) = \delta(\x) \delta(t)
                \quad\Rightarrow\quad
       \delta p(\x,t) = \left[c_0^2 \delta(\x)  - b_0 A^{\gamma-1/2}(\x)\right]\delta(t) - a_0 A^{\gamma-1}(\x) \delta'(t)\,.
$$
Because $\gamma-1/2\in\{1,\,2,\,3,\,\ldots\}$ excludes $\gamma-1\in \{1,\,2,\,3,\,\ldots\}$ and vice versa, property (A) 
implies (for fixed $\gamma>0$ and $t>0$) that for every $n\in\N$ there exists an $\x_n\in\R^3$ with $|\x_n|\geq n$ such that 
$$
     \delta p(\x_n,t)
     \quad \mbox{does not vanish.}
$$ 
From this and the fact that the distributions $\delta(t)$ and $\delta'(t)$ appear in the pressure variation, we infer that  
$\delta p$ is an \emph{instant global} pressure variation caused by a density variation concentrated in a point in space 
and time. Hence the state equation is not local and \emph{if} $\rho$ has a finite wave front speed, then $p$ 
does not have a finite wave front speed.

\section{Density and pressure waves of the Treeby and Cox model}
\label{sec-causality}

In this section we show that the density and pressure waves of the Treeby and Cox model do not have a finite wave front speed. 
For that purpose we start with the derivation of the two governing wave equations.

\subsection*{Two wave equations}

The state equation of Treeby and Cox~(\ref{stateeqTC}) can be written as follows 
\begin{equation}\label{stateeqgen}
     p = c_0^2\,K *_{\x,t} \rho   \qquad \quad \mbox{($*_{\x,t}$ space-time convolution)}\,,  
\end{equation}
with $A^\gamma$ defined as in Eq.~(\ref{defA}) and 
\begin{equation}\label{KTC}
        K(\x,t) := \delta(\x)\,\delta(t) 
                          - \frac{a_0}{c_0^2}\,A^{\gamma-1}(\x)\,\delta'(t) 
                          - \frac{b_0}{c_0^2}\,A^{\gamma-1/2}(\x)\,\delta(t)\,.
\end{equation}
Inserting this state equation into the \emph{linearized Euler and continuity equations}, i.e.
$$
   \rho_0\,\frac{\partial \mathbf{v}}{\partial t} = -\nabla p - \mathbf{f}_0  \qquad\mbox{and}\qquad
   \frac{\partial \rho}{\partial t} + \rho_0\,\nabla\cdot \mathbf{v} = 0\,
$$
yields the wave equations
\begin{equation}\label{waveeqrho}
 K *_{\x,t} \Delta \rho - \frac{1}{c_0^2}\,\frac{\partial^2 \rho}{\partial t^2}
    = -\frac{1}{c_0^2} f\,,  
\end{equation}
and
\begin{equation}\label{waveeqp}
 K *_{\x,t} \Delta p - \frac{1}{c_0^2}\,\frac{\partial^2 p}{\partial t^2}
    = -K *_{\x,t} f\,,
\end{equation}
with $f:=\nabla\cdot \mathbf{f}_0$. 
Conversely, the state equation~(\ref{stateeqgen}) can be inferred from these wave equations. If $K^\dagger$ is such that 
$(K^\dagger *_{\x,t} K)(\x,t) =\delta(\x)\,\delta(t)$, then Eq.~(\ref{waveeqp}) reads as follows 
$$
  \Delta p - \frac{1}{c_0^2}\,K^\dagger *_{\x,t}\frac{\partial^2 p}{\partial t^2}
    = - f\,,
$$
i.e. it is not the same wave equation as in Eq.~(\ref{waveeqrho}). In particular, we see that the modeling of the correct 
source terms of dissipatve waves are crucial.    

The \emph{Green function} of Eq.~(\ref{waveeqrho}) is defined by
\begin{equation}\label{Geq}
    \frac{\partial^2 G}{\partial t^2}
          + a_0\,\frac{\partial}{\partial t} (-\Delta)^\gamma\,G
          + b_0\,(-\Delta)^{\gamma+1/2}\,G
          - c_0^2\,\Delta G
             = \delta(\x)\,\delta(t)
\end{equation}
from which we infer  
$$
    \rho = \frac{1}{c_0^2}\,f *_{\x,t} G   \qquad\mbox{and}\qquad 
    p = f *_{\x,t} K *_{\x,t} G \,.
$$
We see that $\rho$ has a finite wave front speed if and only if $G$ has a finite wave front speed.

\subsection*{Derivation of the Green function of Eq.~(\ref{waveeqrho})}

Fourier transform of Eq.~(\ref{Geq}) with respect to $\x$ yields the \emph{Helmholtz equation}
\begin{equation}\label{Heq}
    \frac{\partial^2 \hat G}{\partial t^2}
          + a_0\,k^\gamma\, \frac{\partial \hat G}{\partial t}
          + (c_0^2\,k^2 + b_0\,k^{\gamma+1/2})\,\hat G
              = \frac{\delta(t)}{(2\,\pi)^{3/2}} \,.
\end{equation}
Here $k:=|\k|$ denotes the \emph{wave number} corresponding to the \emph{wave vector} $\k\in\R^3$. Inserting the ansatz
\begin{equation}\label{ansatz}
       \hat G(\k,t) = [A_1\,e^{-\lambda_1\,t} + A_2\,e^{-\lambda_2\,t}]\,H(t)  \qquad \mbox{($H$ Heaviside function)}
\end{equation}
into the Helmholtz equation leads to the equations
\begin{equation}\label{eqA1A2}
\begin{aligned}
      A_1 + A_2 =0 \,, \qquad\qquad
      (\lambda_1\,A_1 + \lambda_2\,A_2) = - \frac{1}{(2\,\pi)^{3/2}}
\end{aligned}
\end{equation}
and
\begin{equation}\label{eqlambda}
\begin{aligned}
    \lambda_{1,2}^2 - a_0\,k^\gamma\,\lambda_{1,2} + (c_0^2\,k^2 + b_0\,k^{\gamma+1/2})
              = 0 \,.
\end{aligned}
\end{equation}
Hence 
\begin{equation*}
      \hat G(\k,t) 
          = \i\,2\,A_1(k)\,e^{-\mu(k)\,t} \, \sin(\vartheta(k)\,t) \qquad \mbox{for $t>0$}\,,
\end{equation*}
where $\lambda_j = \mu + \i\,(-1)^{j-1}\,\vartheta$ with\footnote{For wave attenuation, it is required that $\mu$ is 
positive and thus the constant $a_0$ has to be positive. Indeed, this is part of our general assumption in 
Eq.~(\ref{Assabc}).}
\begin{equation}\label{muvartheta}
         \mu =  \frac{1}{2}\,a_0\,k^\gamma   \qquad\mbox{and}\qquad
         \vartheta = c_0\,k\,\sqrt{1 - \frac{a_0^2}{4\,c_0^2}\,k^{2\,\gamma-2} + \frac{b_0}{c_0^2}\,k^{\gamma-3/2}}
\end{equation}
for $j=1,\,2$. Solving Eqs.~(\ref{eqA1A2}) with respect to $A_1$ and $A_2$ yields
\begin{equation}\label{solA1A2}
\begin{aligned}
   A_1 
       = \frac{\i}{(2\,\pi)^{3/2}\,2\,\vartheta}
       \qquad\mbox{and}\qquad
   A_2 = - A_1\,
\end{aligned}
\end{equation}
and thus we end up with 
\begin{equation}\label{sol}
      \hat G(\k,t) 
          = \frac{e^{-\mu(k)\,t}}{(2\,\pi)^{3/2}} \, \frac{\sin(\vartheta(k)\,t)}{\vartheta(k)} \,H(t)\,.
\end{equation}
We note that this identity holds for $\vartheta(k)\in\R$ as well as for $\vartheta(k)\in \i\,\R$.

\subsection*{Wave front speed of $G$}

In the Appendix (cf. Examples~\ref{exam:PW02} and~\ref{exam:PW03} and Remark~\ref{rema:pG}), it is shown that the support 
of the Green function $G$ as well as $p := c_0^2\,K *_{\x,t} G$ cannot be bounded, i.e. for every $n\in\N$ 
there exists an $\x_n\in\R^3$ such that
\begin{equation}\label{propGsupp}
     G(\x_n,t) \quad\mbox{does not vanish}   \qquad\mbox{for}\qquad  |\x_n|\geq n  \,,
\end{equation}
where $t>0$ and $\gamma>0$ are fixed. Does this mean that the \emph{wave front speed} of $G$ is not finite? 
Assume that $c_F$ is finite. Then the \emph{travel time} of the wave front traveling from point $\mathbf{0}$ to point $\x$ is
given by
$$
          T(|\x|) = \frac{|\x|}{c_F}
$$
and we expect
$$
        G(\x,t) = 0 \qquad\mbox{for}\qquad  T(|\x|) > t\,.
$$
But, due to property~(\ref{propGsupp}) this does not hold for $\x_n$ with $n>c_F\,t$.This contradiction shows that $c_F<\infty$ 
does not hold, i.e. there exists no finite wave front speed $c_F$ of $G$, $\rho$ and $p$.

\section{Conclusions}

In this paper we showed that the state equation of Treeby and Cox establishes a nonlocal relationship $p=F(\rho)$ 
between the density $\rho$ and the pressure $p$. As a consequence, it follows \emph{if} the density wave $\rho$ 
has a finite wave front speed, then the respective pressure wave $p=F(\rho)$ does not have a finite wave front speed. 
This is in contrast to the state equations for the frequency dependent dissipation models, which satisfy the 
Kramer-Kronig relation. Moreover, we showed that any density wave satisfying 
\begin{itemize}
\item the \emph{state equation} of Treeby and Cox, 

\item the \emph{equation of motion} and 

\item the \emph{equation of continuity}

\end{itemize}
cannot have a finite wave front speed. Similarly, it follows that the respective pressure wave does not have a 
finite wave front speed.

In summary, we have shown that the state equation of the Treeby and Cox violate causality and that the respective 
density and pressure waves violate causality in the sense that they do not have a finite wave front speed. 
Thus it is not an appropriate substitution for those frequency dependent dissipation models 
that obey causality. 

The model might be corrected by modeling a kernel $K$ similarly to Eq.~(\ref{KTC}) which satisfies 
at least the following modification of Eq.~(\ref{KKrel})  
$$
           K(\x,t) = 0 \qquad\mbox{for}\qquad  t<\frac{|\x|}{c_0} \,. 
$$
However, an analysis of the respective wave equation for the density is far more complicated than that one presented 
in this paper. Such an analysis of an appropriate model is beyond the scope of this paper. \\

\section{Appendix: Applications of the Paley-Wiener Theorem}

In this section we focus on the pure mathematical part of this paper that permits to conclude crucial physical statements. 
In essence, we employ the first (and simplest) part of the Paley-Wiener Theorem which is well-known from signal and 
image processing. In what follows, $\hat f$ denotes the Fourier transform of $f$ with respect to the space variable $\x$. 
The Paley-Wiener Theorem reads as follows: 

\begin{theo}[Paley, Wiener]\label{th:PW}
Let $n\in\N$, $|\z|:=\sqrt{\sum_{j=1}^n |z_j|^2}$ for $\z\in\C^n$ and $R>0$. The support of a  distribution $f$ is contained in the 
closed ball $B_R(\mathbf{0})$ if and only if
\begin{itemize}
\item [(PW1)] $\hat f=\hat f(\k)$ ($\k\in\R^n$) can be extended to a holomorphic function $g=g(\z)$ ($\z\in\C^n$) satisfying 
            $\hat f(\z)=g(\z)$ and 

\item [(PW2)] there exist constants $C>0$ and $N>0$ such that
      $$
         |g(\z)| \leq C\,(1+|\z|)^N\,e^{R\,|\Im(\z)|}  \qquad \mbox{for} \qquad \z\in\C^3\,. 
      $$
\end{itemize}
\end{theo}

The previous Theorem states that if a function or distribution $f$ is 
such that $g:\z\mapsto \hat f(\z)$ is not holomorphic, then $\hat f$ cannot have a bounded support in the closed ball 
$B_R(\mathbf{0})$ for any $R>0$. $g$ is not holomorphic means that at least one of the following properties are 
not satisfied:
\begin{itemize}
\item [(H1)] the function $z_j\mapsto g(z_1,\ldots z_j,\,\ldots z_n)$ is \emph{holomorphic} for every $j\in \{1,\,2,\,\ldots,\,n\}$  and 

\item [(H2)] $g$ is \emph{locally bounded}, i.e. for each point $\w\in\C^n$ there exist constants $M>0$ and $\epsilon>0$ such that 
             $|g(\z)|\leq M$ for all $\z\in B_\epsilon(\w)$.
\end{itemize}
In what follows, these facts are used to show the nonlocality of the state equation of Treeby and Cox and 
the infinite speed of the solutions of the density and pressure wave equation of Treeby and Cox.

\begin{rema}
For models of wave dissipation, the second condition of the Paley-Wiener Theorem is usually satisfied for vectors of real 
numbers, because $\hat f:\R\to\R$ decreases exponentially. However, the significance of this condition is due to its 
validity for all vectors of complex numbers. Frequently, $f$ is not contained in a closed ball of finite radius due to 
the fact that $\hat f$ is holomorphic, but there exists a sequence of complex vectors $\z_n\in\C^3$ for $n\in\N$ 
such that $(|g(\z_n)|)_{n\in\N}$ does not \emph{decrease exponentially} for $n\to\infty$, but \emph{increases} much 
faster than $(e^{C\,|\z_n|})_{n\in\N}$, where $C>0$ is a constant. For example, it may increase like 
$(e^{C\,|\z_n|^2})_{n\in\N}$. 
%We see that complex analysis plays a \emph{key role} for every analysis of causality. 
\end{rema}

\subsection*{Nonlocality of $(-\Delta)^{d/2}$ for $d>0$ with $d\not\in\{2,\,4,\,6,\,\ldots\}$}

In our first example, we exploit the fact that 
$$
      \k\in\R^3 \mapsto |\k|^d   \qquad\mbox{is not infinitely differentiable at zero,} 
$$
if $d>0$ with $d\not\in \{2,\,4,\,6,\,\ldots\}$ and show property (A) stated in Section~\ref{sec-stateeq}. 
For example, consider the function $f(\k):=|\k|^3$ for which we have 
$$
  \frac{\partial^2 f}{\partial k_1^2}(\k) 
     = \frac{3\,k_1^2}{|\k|} + 3\,|\k|\,.
$$ 
Due to the rule of de l'Hospital, we see that this function is everywhere continuous, however, the second term has an edge 
at $\k=\mathbf{0}$ and thus it is not differentiable with respect to $k_1$ at zero. Moreover, for $d=2\,m$ with 
$m \in\N$, we have $|\k|^d = (k_1^2+k_2^2+k_3^2)^m$ and thus $\k\in\R^3 \mapsto |\k|^{2\,m}$ is infinitely differentiable.

\begin{exam}\label{exam:PW01}
Property (PW1) is not satisfied for
\begin{equation}\label{deff1}
         f := (-\Delta)^{d/2} \delta(\x)   \qquad\mbox{if}\qquad  d>0 \quad\mbox{with}\quad 
                                                    d\not\in\{2,\,4,\,6,\,\ldots\}\\,.
\end{equation}
Indeed, Fourier transform of $f$ yields
$$
       \hat f(\k) = [-(-\i\,|\k|)^2]^{d/2}\,(2\,\pi)^{-3/2} = (2\,\pi)^{-3/2}\,|\k|^d \,,
$$
where $k_1\in\R \mapsto |k_1|^d$ is not infinitely differentiable (for $d>0$ and $d\not\in\{2,\,4,\,6,\,\ldots\}$) 
and thus $z_1\in\C \mapsto |z_1|^d$ is not holomorphic. Because (H1) does not hold, it follows that 
$\z\in\C^3 \mapsto |\z|^d$ is not holomorphic and consequently $\hat f$ is not holomorphic, too. This shows that property 
(PW1) of the Paley-Wiener Theorem is not satisfied for $f$ defined as in Eq.~(\ref{deff1}) and hence the support of $f$ 
cannot lie in an open ball of finite radius. 
In other words, for every $n\in\N$ there exists an $\x_n\in\R^3$ with $|\x_n|\geq n$ such that
\begin{equation}\label{vanish}
         f(\x_n)   \qquad \mbox{does not vanish.}
\end{equation}
\end{exam}

\begin{rema}
What do we mean by Eq.~(\ref{vanish}), if $f$ is a distribution?  
It means that for arbitrary small $\epsilon>0$, the restriction of the distribution $f$ to the set $B_{\epsilon}(\x_n)$ 
is not the zero function (=zero distribution), where $B_{\epsilon}(\x_n)$ denotes the open ball with center $\x_n$ and 
radius $\epsilon$. In other words, the integral of $f$ over $B_{\epsilon}(\x_n)$ does not vanish for arbitrary small 
$\epsilon>0$. 
\end{rema}

\subsection*{Unbounded support of $G$}

Let $f$ be a distribution with a Fourier transform of the form 
\begin{equation}\label{strucG}
           \hat f(\k) = g(\theta(\k))\,h(\k) \,, 
\end{equation}
where $g,\,h:\R\to\R$ are well-defined and nice functions.  
From the \emph{product} and \emph{chain rule} of differentiation, it follows that $\hat f$ is not $n$-times differentiable 
at zero if 
\begin{itemize}
\item [a)] $\theta(\k)$ is not $n$-times differentiable at $\k=\mathbf{0}$ and 
             
\item [b)] $h(\mathbf{0})\not=0$ and $g'(\theta(\mathbf{0}))\not=0$.          
\end{itemize}
For example, 
$$
   \frac{\partial^2 \hat f}{\partial k_1^2}(\mathbf{0}) 
     =  \frac{\partial^2 \theta}{\partial k_1^2}(\mathbf{0})\,g'(\theta(\mathbf{0}))\,h(\mathbf{0})  + \cdots  \,,
$$
i.e. $\hat f$ is not two times differentiable at zero if $\theta$ is not two times differentiable at zero, 
$h(\mathbf{0})\not=0$ and $g'(\theta(\mathbf{0}))\not=0$.

\begin{exam}\label{exam:PW02}
Let $T>0$, $\gamma>0$ with $\gamma\not\in \{2,\,4,\,6,\,\ldots\}$ and the function $f$ be such that its Fourier transform is 
as in Eq.~(\ref{strucG}) with 
$$
       g(s) = e^{-s\,T}\,,   \qquad 
        \theta(\k) = |\k|^\gamma   \qquad\mbox{and}\qquad 
       h(\mathbf{0}) \not= 0\,.
$$
Because $g'(\mathbf{0})\not=0$ and $\theta$ is not $n$-times differentiable for $n\geq \gamma$ with $n\in\N$, 
it follows that $\hat f$ is not $n$-times differentiable. Hence, $\hat f$ is not holomorphic on $\C^3$ and thus, due to 
the Paley-Wiener Theorem, the distribution $f$ does not have a bounded support. 
In other words, for every $n\in\N$ there exists an $\x_n\in\R^3$ with $|\x_n|\geq n$ such that
$$
         f(\x_n)   \qquad \mbox{does not vanish.}
$$ 
\end{exam}

\vspace{0.5cm}
The Green function of the density wave equation of Treeby and Cox (cf. Eqs.~(\ref{sol}) and~(\ref{muvartheta})) has the same 
form as in Example~\ref{exam:PW02}, where 
$$
       h(\k) = \frac{\sin(\vartheta(|\k|)\,t)}{\vartheta(|\k|)}\,\frac{H(t)}{(2\,\pi)^{3/2}}  
$$
with
$$
       \vartheta = c_0\,\sqrt{|\k|^2 - \frac{a_0^2}{4\,c_0^2}\,|\k|^{2\,\gamma} + \frac{b_0}{c_0^2}\,|\k|^{\gamma+1/2}}\,.
$$
Because $h(\mathbf{0}) \not= 0$ holds for $t>0$, Example~\ref{exam:PW02} implies that for every $n\in\N$ there exists 
an $\x_n\in\R^3$ with $|\x_n|\geq n$ such that
$$
         G(\x_n,t)   \qquad \mbox{does not vanish,}
$$ 
for fixed $t>0$ and $\gamma>0$ with $\gamma\not\in \{2,\,4,\,6,\,\ldots\}$. The remaining cases 
$\gamma\in \{2,\,4,\,6,\,\ldots\}$ follow from the next example. For this example, we require the second part of the 
Paley-Wiener Theorem.

\begin{exam}\label{exam:PW03}
Let $f$ be as in Example~\ref{exam:PW02} with $\gamma\in \{2,\,4,\,6,\,\ldots\}$ and  
\begin{equation}\label{proph}
        h(\i^{2/\gamma}\,k_1,0,0) \geq D >0 \qquad \mbox{for}\qquad  k_1 \to \infty\,
\end{equation}
hold for some constant $D>0$. Then $f$ does not have a bounded support. We assume that $f$ has bounded support and show 
a contradiction. For $d:=\gamma/2$ we have  
$$
          |\k|^\gamma = \left(k_1^2 +k_2^2+k_3^2\right)^d
$$
and consequently $\z\in\C^3\mapsto |\k|^\gamma$ is holomorphic. Moreover, for $\z=(\i^{1/d}\,k_1,0,0)$ with $k_1>1$, we have 
$$
     |\z|^\gamma = \left( \i^{1/d}\,k_1\right)^{2\,d} =  - k_1^\gamma \, \leq  \,-k_1^2 \,\stackrel{!}{<}\, 0 \,
$$
and
$$
            0 <  \Im(\i^{2/\gamma})\,k_1 \leq k_1    \qquad\mbox{for}\qquad   k_1>0  \,.
$$
Hence, for this choice of $\z$, condition (PW2) of the Paley-Wiener Theorem with property~(\ref{proph}) 
implies that there exist constants $C>0$ and $N>0$ such that\footnote{The crucial point is that the left hand side is not 
decreasing but increasing for an appropriate sequence of complex numbers.}
$$
         D\,\exp(T\,k_1^2) \leq C\,(1+k_1)^N\,\exp(R\,k_1)    \qquad \mbox{for} \qquad |k_1| \to \infty\,.  
$$ 
But this inequality cannot hold for positive constants $D$, $T$, $C$, $N$ and $R$. 
This contradiction shows that the support of $f$ cannot lie in an open ball of finite radius $R$. 
\end{exam}

\vspace{0.5cm}
Let $\gamma\in \{2,\,4,\,6,\,\ldots\}$ and $\k=(\i^{2/\gamma}\,k_1,0,0)$ for $k_1>1$. 
For the Green function $G$ of the density wave equation of Treeby and Cox, we have 
$$
   h := \frac{\sin(\vartheta\,t)}{\vartheta}    \qquad\mbox{and}\qquad 
   \vartheta(\i^{2/\gamma}\,\k) \approx  \i\,\frac{a_0}{2}\,k_1^\gamma \qquad\mbox{for large $k_1>1$}
$$
and thus, due to $\sin(\i\,x)=\i\,\sinh(x)$, 
$$
     h(\i^{2/\gamma}\,\k) 
          \approx  \frac{2\,\sinh(\frac{a_0}{2}\,t\,k_1^\gamma) }{a_0\,k_1^\gamma} \qquad\mbox{for large $k_1>0$.}
$$
But his means that property~(\ref{proph}) holds and therefore the support of $G$ cannot lie in an open ball of 
finite radius due to Example~\ref{exam:PW03}.

\begin{rema}\label{rema:pG}
The above examples show that the Green function of the density wave of Treeby and Cox does not have a finite wave front speed, 
but what does this mean for the respective pressure wave $p_G := c_0^2\,(K_1 *_{\x,t} G)$? 

Let $\gamma>0$ with $\gamma\not\in \{2,\,4,\,6,\,\ldots\}$. Then it is easy to see that 
$$
   \hat p_G(\k,t) 
      =   c_0^2\,\hat G(\k,t)
       - a_0 \,  |\k|^{2\,(\gamma-1)}\,\frac{\partial \hat G}{\partial t} (\k,t)
       - b_0\, |\k|^{2\,\gamma-1}\,\hat G(\k,t)
$$
is not infinitely differentiable, because $\hat G(\k,t)$ is not infinitely differentiable.
If $\gamma\in \{2,\,4,\,6,\,\ldots\}$, then property (PW2) of the Paley-Wiener Theorem 
does not hold, because 
$$
    |\hat p_G(\i^{2/\gamma}\,k_1,0,0,t) |  \qquad\mbox{grows like}\qquad 
    \exp(C\,k_1^2)  \qquad\mbox{for $k_1\to\infty$,}
$$
where $C>0$ is a constant.  
As in Examples~\ref{exam:PW02} and~\ref{exam:PW03}, it follows  that for every $t>0$ the support of 
$p_G(\cdot,t)$ cannot lie in an open ball of finite radius.  
\end{rema}

\end{document}